\documentclass[conference,a4paper]{APSIPA2026}
\usepackage{amsmath}
\usepackage{graphicx}
\usepackage{multirow}
\usepackage{threeparttable}
\usepackage[
  backend=biber,
  style=ieee,
  doi=false,
  url=false,
  isbn=false, 
  eprint=true,
  giveninits=true
]{biblatex}
\AtEveryBibitem{%
  \clearfield{issn}%
  \clearfield{isbn}%
  \clearfield{url}%
  \clearfield{urldate}%
  \clearfield{abstract}%
  \clearfield{keywords}%
  \clearfield{lccn}%
  \clearfield{series}%
}
\usepackage{geometry}
\usepackage{fancyhdr}

\fancypagestyle{firststyle}{
  \fancyhf{}
}

\usepackage{amssymb,bm} 

\begin{document}

\title{Feedforward Active Speech Suppression Based on Time Series Prediction of Speech Signals Using Neural Networks}

\author{
\authorblockN{
Manami Nishikata\authorrefmark{1} and
Shoichi Koyama\authorrefmark{2}\authorrefmark{1}
}
\authorblockA{
\authorrefmark{1}
Graduate Institute for Advanced Studies, SOKENDAI, Kanagawa, Japan \\
E-mail: m-nishikata@nii.ac.jp
}
\authorblockA{
\authorrefmark{2}
National Institute of Informatics, Tokyo, Japan \\
E-mail: koyama.shoichi@ieee.org
}
}

\maketitle
\thispagestyle{firststyle}
\pagestyle{empty}

\begin{abstract}
A feedforward active noise control (ANC) method based on time-series prediction for speech signals is proposed. 
Although current ANC techniques are highly effective against stationary noise, suppressing highly non-stationary speech signals remains a challenging task. 
We propose an adaptive filtering algorithm for active speech suppression based on neural-network-based time-series prediction of future signals.
The update value for the linear control filter is calculated based on the predicted signal, as well as the current and past signals. 
Numerical experiments indicated that the noise reduction can be improved in both cases: when using the true predicted signal and when using a signal predicted by neural networks. 
\end{abstract}

\section{Introduction}
Active noise control (ANC) is a technique that reduces incoming noise at a target position by emitting anti-noise through a loudspeaker~\cite{linearanc,anc1975}, which has been applied in headphones, vehicle cabins, and air ducts, for example~\cite{duct1998,car,car2,cabin2022}. In the feedforward ANC techniques, which enable the suppression of broadband noise, a control filter is applied to the signal from the primary noise source captured by the reference microphone to generate the signal of the secondary loudspeaker. 
The control filter is adaptively updated using adaptive filtering algorithms to suppress various types of noise by monitoring the noise status at the target position with the error microphone. 

In conventional feedforward ANC, the primary target for suppression has been (quasi-)stationary noise, such as air-conditioning, engine, and duct noise~\cite{ancfffb}. However, in office environments, for example, there may be situations where it is desirable to suppress speech signals. Since speech signals are highly non-stationary, with drastic fluctuations in amplitude, spectrum, and activity over short periods of time, it is difficult to track these using adaptive filters~\cite{chu2003speech,deng2003speech,staticspeech}. For example, in the filtered-x least-mean-square (FxLMS) algorithm~\cite{Fxlms,winneradapt,adapt1981}, the filter is updated to minimize the expected power of the error signal; however, because the update values are approximated using instantaneous correlations of the observed signals, its tracking performance is not sufficient for speech signals. Although several methods for determining filter update values using past observation signals, e.g., recursive least squares (RLS)~\cite{RLS} and affine projection method~\cite{fastapaanc}, are known, they incur high computational cost due to the computation of inverse matrices.

We propose an active speech suppression method based on time-series prediction of speech. Using rapidly advancing neural network (NN) techniques for speech generation, future speech signals are predicted and used to update the adaptive filters in FxLMS. There are studies on linear prediction of speech signals for delay compensation in ANC~\cite{Iotov,speechlinearpred} and on NN-based control-signal generation, mainly for nonlinear ANC systems~\cite{nonlinearanc,deepanc,dnoisenet,deepasc,rnn2023}. Our contribution lies in 1) formulating the update of linear filters using speech time-series prediction within a framework similar to that of FxLMS, and 2) in evaluating its performance using true predicted signals and signals predicted by a simple NN. 
\section{Problem statement and prior work}
A secondary loudspeaker emits an anti-noise signal $y_k$ at the discrete time $k$ to suppress primary noise at the error microphone's position $d_k$ as
\begin{align}
e_k = d_k + \sum_{l=0}^{L-1} s_l y_{k-l},
\end{align}
where $s_l$ ($l=0, \ldots, L-1$) is the $L$-length secondary path from the secondary loudspeaker to the error microphone, and $e_k$ is the error microphone signal. $d_k$ is also referred to as the desired signal. In the feedforward ANC, $y_k$ is obtained by convolving the $K$-length control filter $\bm{w}_k=[w_0, \ldots, w_{K-1}]^{\mathsf{T}}$ with the reference signal as
\begin{align}
y_k = \bm{w}_k^{\mathsf{T}} \bm{x}_{k}, 
\end{align}
where $x_k$ is the signal of the reference microphone placed near the primary noise source, and $\bm{x}_{k}=[x_{k}, \ldots, x_{k-K+1}]^{\mathsf{T}}$.

The control filter is adaptively obtained based on the cost function defined as the expected squared power of the error signal, as
\begin{align}
\mathcal{J} = \mathbb{E}[|e_k|^2],
\end{align}
where $\mathbb{E}[\cdot]$ denotes the expectation operation. Thus, $\bm{w}_k$ is updated to minimize $\mathcal{J}$ as
\begin{align}
\bm{w}_{k+1} &= \bm{w}_k - \mu_k \mathbb{E}[e_k\bm{x}_{\mathrm{f},k}],\\
x_{\mathrm{f},k} &= \sum_{l=0}^{L-1} \hat{s}_l x_{k-l}.
\end{align}
Here, $x_{\mathrm{f},k}$ is the filtered reference signal, and $\bm{x}_{\mathrm{f},k}=[x_{\mathrm{f},k}, \ldots, x_{\mathrm{f},k-K+1}]^{\mathsf{T}}$. We assume that the estimated secondary path $\hat{s}_l$ is given in advance. 
The normalized step size parameter $\mu_k$ is updated as
\begin{align}
\mu_k = \frac{\mu_0}{\|\bm{x}_{\mathrm{f},k}\|^2 + \epsilon},
\label{eq:stepseze}
\end{align}
where $\mu_0$ is the step size parameter, and $\epsilon$ is the regularization parameter. In the FxLMS algorithm, the expectation operation is replaced with the instantaneous value as
\begin{align}
\bm{w}_{k+1} = \bm{w}_k - \mu_k e_k\bm{x}_{\mathrm{f},k}.
\label{eq:update_fxlms}
\end{align}

When the primary noise signal is stationary, the control filter obtained using FxLMS is expected to converge stochastically to the optimal filter given by the Wiener filter, with an appropriate step size parameter $\mu_0$~\cite{winneradapt}. When the primary noise is a non-stationary signal, such as speech, the signal's statistical properties vary over time; consequently, the optimal filter changes constantly, making it difficult for FxLMS to follow. 

Several methods have been proposed to improve the tracking performance of adaptive filters by using not only instantaneous values but also past observation signals in gradient computations. A well-known technique is the RLS algorithm~\cite{RLS,lmsadapt}, which sequentially computes the least-squares solution for past observation signals weighted by a forgetting factor. However, RLS requires the sequential updating of the inverse matrix, resulting in a high computational cost and instability due to the accumulation of numerical errors. 
Affine projection algorithms~\cite{fastapaanc} update the filter using multiple recent observations through a projection operation, which generally requires matrix inversion depending on the projection order.

In prior work on ANC for speech, a delay-compensation method based on linear prediction of the speech signal has been proposed to address the causality constraints inherent in ANC headphones~\cite{Iotov,speechlinearpred}. Predictive fixed-filter ANC (PFANC)~\cite{pfanc} predicts next-frame control filters using a convolutional recurrent NN. Several methods that directly generate control signals from speech or non-stationary signals using NNs have also been proposed, especially for nonlinear systems~\cite{deepanc,dnoisenet,deepasc}.

\section{Speech-prediction-based ANC}

We propose an ANC method for suppressing speech based on predicting future time-series signals. The FxLMS algorithm is extended to incorporate predicted speech time series obtained using NNs, which is referred to as the \textit{speech-prediction-based FxLMS algorithm} (SP-FxLMS).  

\begin{figure}[t!]
\centering
\includegraphics[width=0.95\linewidth]{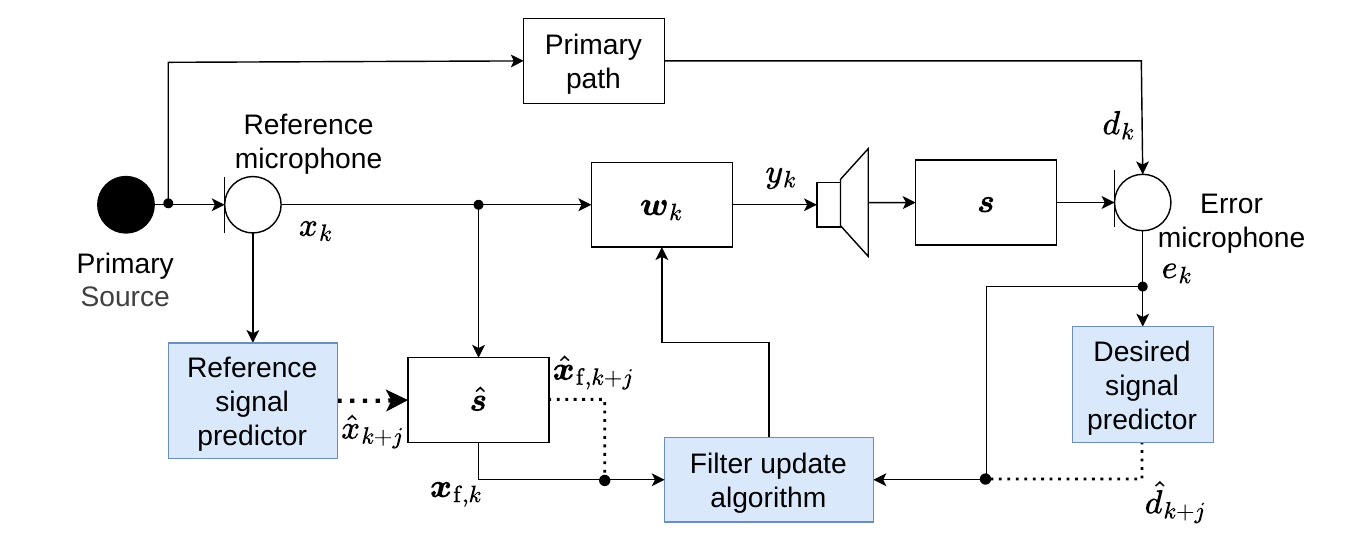}
\caption{
Block diagram of the proposed SP-FxLMS. The time-series prediction of the reference and desired signals is performed by the predictors, and the predicted signals are used for updating the control filter $\bm{w}_k$. 
}
\label{fig:ancblock}
\end{figure}

\subsection{Speech-prediction-based FxLMS algorithm}

First, we formulate the SP-FxLMS algorithm, which updates the control filter using the predicted speech time series to improve the tracking performance for non-stationary signals. The schematic diagram of SP-FxLMS is shown in Fig.~\ref{fig:ancblock}.

The reference and desired signal predictors output the reference and desired signals predicted up to $a$ samples 
\begin{align}
\hat{x}_{k+1}, \ldots, \hat{x}_{k+a} \\
\hat{d}_{k+1}, \ldots, \hat{d}_{k+a}
\end{align}
from the observed signals up to time $k$, respectively. Thus, the filtered reference signal for the $j$-sample predicted reference signal ($j=1,\ldots,a$) is written as $\hat{\bm{x}}_{\mathrm{f},k+j}=[\hat{x}_{\mathrm{f},k+j}, \ldots, \hat{x}_{\mathrm{f},k+j-K+1}]$. The $j$-sample predicted error signal is expressed as
\begin{align}
\hat{e}_{k+j} = \hat{d}_{k+j} + \bm{w}_{k}^{\mathsf{T}} \hat{\bm{x}}_{\mathrm{f},k+j}.
\end{align}
We assume the $j$-sample predicted reference and desired signals are obtained by a time-series prediction method. 

\subsubsection{Gradient using future and past signals}

In the standard FxLMS algorithm, the gradient of the cost function at time $k$ is obtained as
\begin{equation}
\bm{\Delta}_k^{\mathrm{C}} = \mu_k e_k \bm{x}_{\mathrm{f},k}.
\end{equation}
We define the gradient for the predicted future signals at time $k$ as
\begin{align}
\bm{\Delta}_k^{\mathrm{F}} = \sum_{j=1}^a \gamma_j^{\mathrm{F}} \mu_{k+j} \hat{e}_{k+j} \hat{\bm{x}}_{\mathrm{f},k+j},
\end{align}
where $\gamma_j^{\mathrm{F}}$ is the weight coefficient obtained by using the forgetting factor $\lambda$ as
\begin{equation}
    \gamma_j^{\mathrm{F}} = \frac{\lambda^j}{\sum_{i=1}^{a} \lambda^i}.
    \label{eq:weight_future}
\end{equation}

We also consider using past reference and desired signals to compute the gradient. When $b$-sample past signals are available, the past error signals are obtained as
\begin{equation}
e_{k-j} = d_{k-j} + \bm{w}_k^{\mathsf{T}} \bm{x}_{\mathrm{f},k-j}.
\end{equation}
Thus, the gradient for the past signals at time $k$ is defined as
\begin{align}
\bm{\Delta}_k^{\mathrm{P}} = \sum_{j=1}^b \gamma_j^{\mathrm{P}} \mu_{k-j} e_{k-j} \bm{x}_{\mathrm{f},k-j},
\end{align}
where 
\begin{equation}
    \gamma_j^{\mathrm{P}} = \frac{\lambda^j}{\sum_{i=1}^{b} \lambda^i}.
    \label{eq:weight_past}
\end{equation}
Note that $\bm{\Delta}_k^{\mathrm{P}}$ is obtained by the observed signals. 

The control filter is updated by combining the gradient for the current, future, and past signals as
\begin{align}
\bm{w}_{k+1} = \bm{w}_k - \frac{1}{|\mathcal{T}|} \sum_{\mathrm{T}\in\mathcal{T}} \bm{\Delta}_k^{\mathrm{T}}, \quad (\mathcal{T} \subseteq \{\mathrm{C},\mathrm{F},\mathrm{P}\})
\label{eq:update_sp-fxlms}
\end{align}
It is possible to use only the current and future signals $\mathcal{T}=\{\mathrm{C},\mathrm{F}\}$ and to use the current, future, and past signals $\mathcal{T}=\{\mathrm{C},\mathrm{F},\mathrm{P}\}$.

Excluding time-series prediction, the filter update in Eq.~\eqref{eq:update_sp-fxlms} has complexity $O((K+a+b)L+(a+b)K)$ for $\mathcal{T}=\{\mathrm{C},\mathrm{F},\mathrm{P}\}$,
compared with $O(KL+K)$ for standard FxLMS and $O(K^2+KL)$ for RLS.

\subsubsection{Variable forgetting factor}

In Eqs.~\eqref{eq:weight_future}, \eqref{eq:weight_past}, the forgetting factor $\lambda$ to obtain the weight coefficients $\gamma_j^{\mathrm{F}}$ and $\gamma_j^{\mathrm{P}}$ is assumed to be a fixed constant. However, it is also possible to make the forgetting factor adaptive to the power of the reference signal, which is referred to as the \textit{variable forgetting factor} and has been applied in adaptive filtering algorithms, such as RLS~\cite{FORTESCUE1981831,VFF2008,albu2012ivff}. Since speech signals normally contain voiced and unvoiced segments, the forgetting factor is set larger when the input power is low and smaller when it is high to improve tracking performance. 

The variable forgetting factor at time $k$ is denoted by $\lambda_k$ and its minimum and maximum values are defined as $\lambda_{\min}$ and $\lambda_{\max}$, respectively ($0 < \lambda_{\min} < \lambda_{\max} \le 1$). The moving average of the power of the filtered reference signal vector $\bm{x}_{\mathrm{f},k}$ at time $k$ is defined as
\begin{align}
\rho_k = \alpha \rho_{k-1} + (1-\alpha) \frac{\|\bm{x}_{\mathrm{f},k}\|^2}{K},
\end{align}
where $\alpha$ ($0 \le \alpha < 1$) is the smoothing coefficients. Thus, $\lambda_k$ is calculated as
\begin{equation}
\lambda_k = \lambda_{\max} - (\lambda_{\max}-\lambda_{\min})\frac{\rho_k}{\rho_k+\rho_0},
\end{equation}
with $\rho_0=\|\bm{x}_{\mathrm{f},k}\|^2/K$.

\subsection{Neural-network-based speech prediction}

We apply a multilayer perceptron (MLP)~\cite{mlp} as a predictor of the speech time series. Although MLP is not expected to achieve high prediction accuracy, it can be implemented with a simple structure. The input is the $N$-sample past and current reference/desired signals normalized by the maximum absolute value, and the output is the $a$-sample predictions. The desired signal can be obtained as $d_k=e_k-\sum_{l} \hat{s}_l y_{k-l}$.
The NN weights are optimized to minimize the following loss function between the predicted and true values:
\begin{equation}
\mathcal{L} =  \frac{1}{a} \sum_{j=1}^{a} \left(\hat{z}_{k+j} - z_{k+j} \right)^2,
\end{equation}
where $\hat{z}_k$ and $z_k$ are the predicted and true signals of $x_k$ or $d_k$.

\section{Experiments}

We conducted two numerical experiments in the free field: 1) experiments using the true predicted signals to evaluate the effect of using the time-series prediction in FxLMS, 2) experiments using the signals predicted by NN to evaluate the proposed SP-FxLMS, including the prediction error. 

\subsection{Experimental setting}

The distance between the secondary loudspeaker and error microphone (secondary path) was $1.0~\mathrm{m}$, and that between the primary source and error microphone (primary path) was $1.5~\mathrm{m}$. The reference signal was directly obtained from the primary source. 
The sound speed was set to $343~\mathrm{m/s}$.
The speech signal for the primary noise source was taken from LibriSpeech ASR corpus (dev-clean)~\cite{libri}. The sampling frequency was $16~\mathrm{kHz}$. 
White Gaussian noise was added to the error signal so that the signal-to-noise ratio (SNR) becomes $30~\mathrm{dB}$.

We compare four types of the proposed SP-FxLMS algorithm with two baselines as follows:
\begin{itemize}
\item[] \textbf{SP-FxLMS\_F:} The proposed SP-FxLMS algorithm using $\mathcal{T}=\{\mathrm{C},\mathrm{F}\}$ in Eq.~\eqref{eq:update_sp-fxlms} with fixed forgetting factor
\item[] \textbf{SP-FxLMS\_F+P:} The proposed SP-FxLMS algorithm using $\mathcal{T}=\{\mathrm{C},\mathrm{F},\mathrm{P}\}$ in Eq.~\eqref{eq:update_sp-fxlms} with fixed forgetting factor
\item[] \textbf{SP-FxLMS\_F\_VFF:} SP-FxLMS\_F with variable forgetting factor
\item[] \textbf{SP-FxLMS\_F+P\_VFF:} SP-FxLMS\_F+P with variable forgetting factor
\item[] \textbf{FxLMS:} The standard FxLMS (Eq.~\eqref{eq:update_fxlms})
\item[] \textbf{FxLMS\_P:} FxLMS using past observations, i.e., $\mathcal{T}=\{\mathrm{C}, \mathrm{P}\}$ in Eq.~\eqref{eq:update_sp-fxlms} with fixed forgetting factor
\end{itemize}

The length of the control filter $\bm{w}_k$ was $K=192$. 
Based on preliminary experiments, a common step size of $\mu_0=1$ was selected for all methods from $\{0.1,0.5,1.0,2.0\}$. Note that, therefore, the step size parameter was not optimized for each method. 
The prediction length in SP-FxLMS was $a=2^n$ ($n\in[0,11]$).
The past length $b$ was set to $1024$.
Unless otherwise specified, the fixed forgetting factor was set to $\lambda=1.0$, i.e., equally weighted for past and predicted signals. 
The variable forgetting factor was set by $\lambda_{\min}=0.9$ and $\lambda_{\max}=0.999$.

For the evaluation measure of the speech reduction, we define the normalized power reduction, $\mathrm{NPR}$, as
\begin{equation}
\mathrm{NPR} = 10\log_{10}\left(\frac{\sum_{k\in\mathcal{I}}\tilde{e}_k^2}{\sum_{k\in\mathcal{I}}d_k^2}\right),
\end{equation}
where $\tilde{e}_k$ is the error signal without the sensor noise and $\mathcal{I}$ denotes the average interval.  
The duration of the primary speech signal was $12~\mathrm{s}$, by concatenating or truncating single speaker's speech signals. The first $5~\mathrm{s}$ was for the adaptation of the filter, and the last $7~\mathrm{s}$ was for the evaluation. 
We define the average $\mathrm{NPR}$ as the $\mathrm{NPR}$ calculated over the entire evaluation interval. Alternatively, we define the frame-wise $\mathrm{NPR}$ as the $\mathrm{NPR}$ averaged for each frame of $1600~\mathrm{samples}$ with a shift of $800~\mathrm{samples}$. 

\subsection{Experiments using true predicted signals}

\begin{figure}[t!]
\centering
\includegraphics[width=0.90\linewidth]{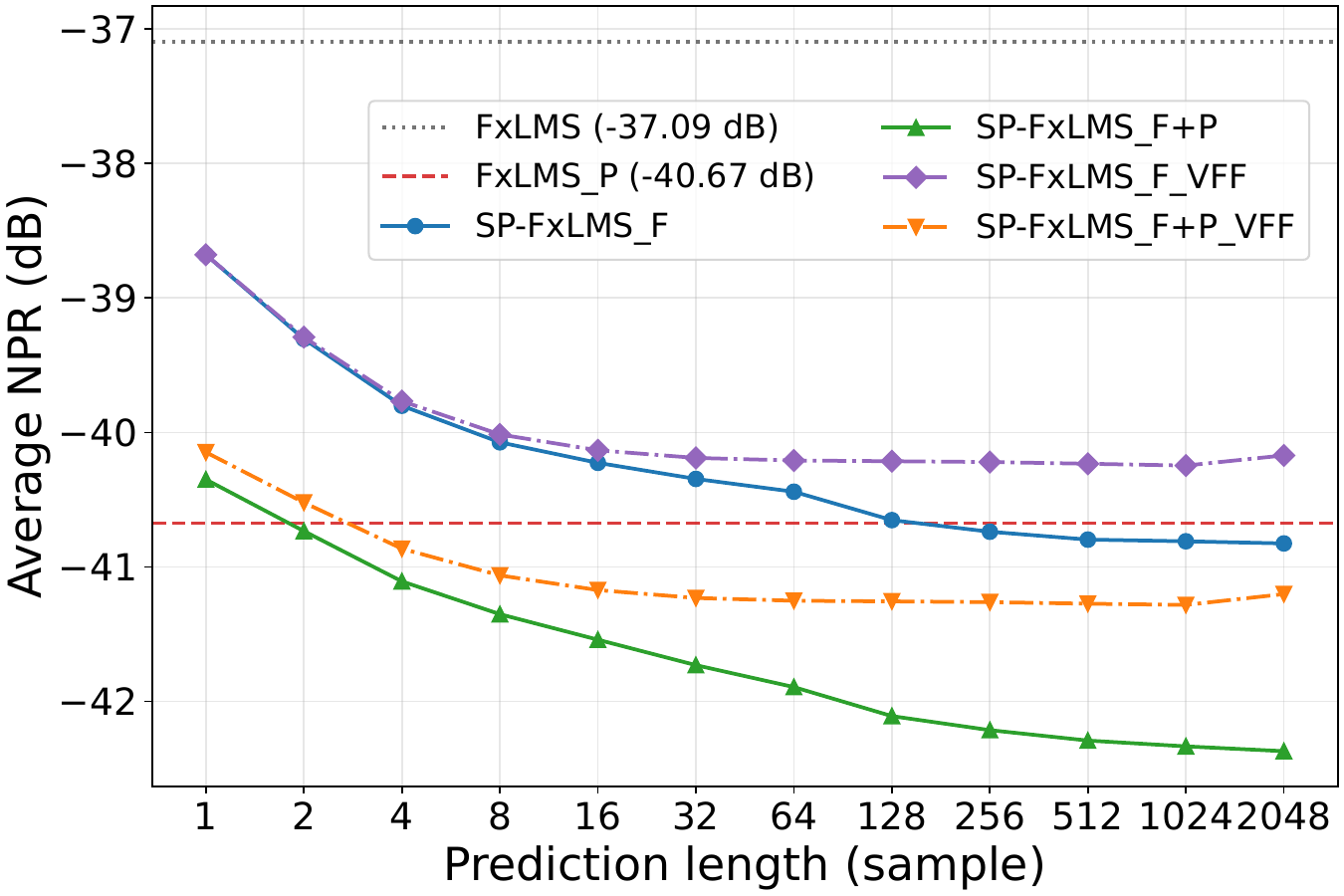}
\caption{
Average $\mathrm{NPR}$ with respect to the prediction length when using the true predicted signals. 
}
\label{1988oracle}
\end{figure}

\begin{figure}[t!]
\centering
\includegraphics[width=0.90\linewidth]{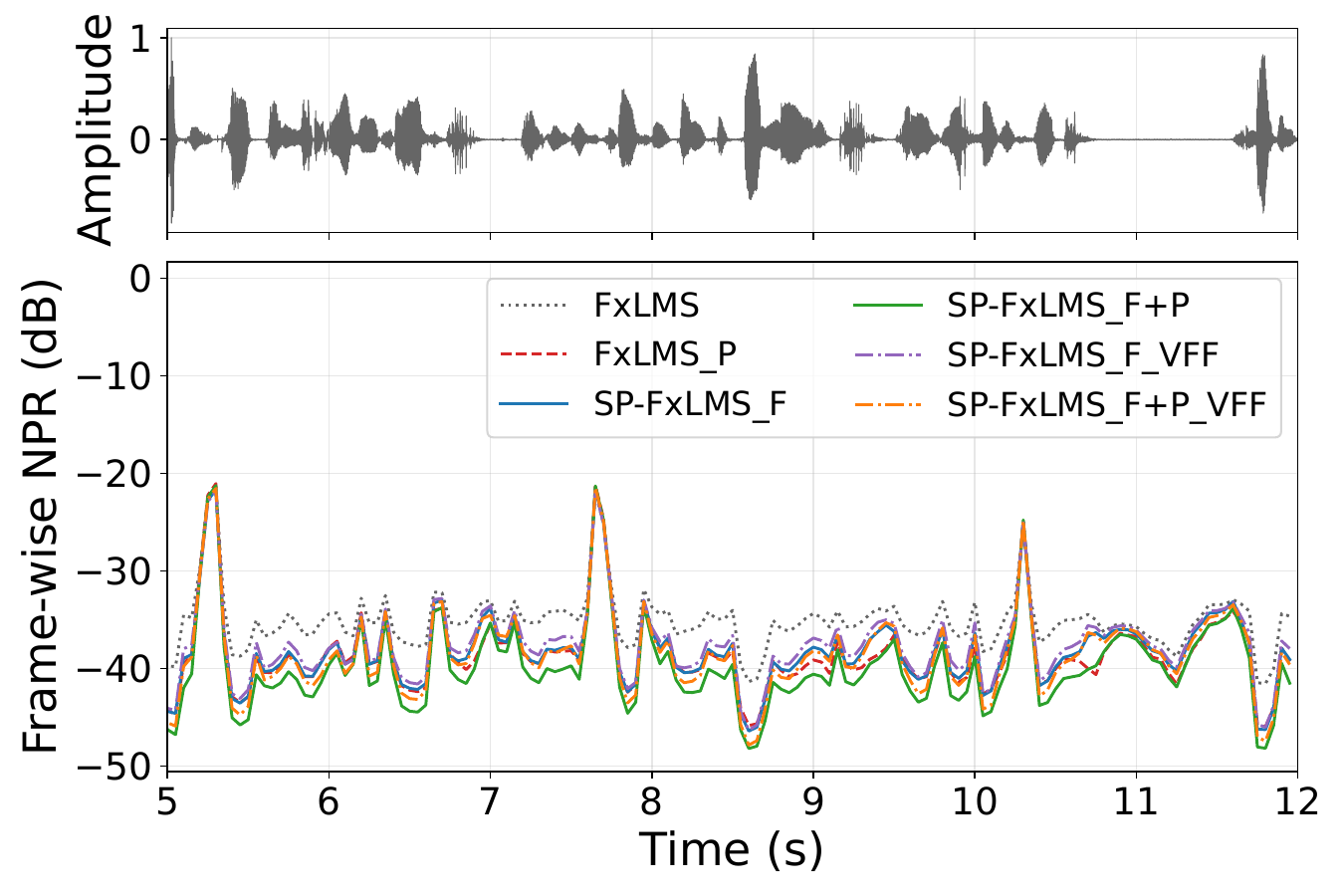}
\caption{
Waveform of the reference signal and frame-wise $\mathrm{NPR}$ when using the true predicted signals for the prediction length $a=128$.
}
\label{1988oracletime}
\end{figure}

First, the case in which the true predicted signal is used is investigated. 
Figure~\ref{1988oracle} shows the average $\mathrm{NPR}$ with respect to the prediction length for \texttt{speaker 1988}. 
Compared to FxLMS, SP-FxLMS and FxLMS\_P achieved lower average NPRs, demonstrating that the gradient computation using past and/or future signals is effective. 
The benefit of using a predicted signal becomes apparent when past signals are also used and when the prediction length is larger than 2 samples. 
Although the average NPR generally decreased as the prediction length increased, it leveled off at around 128 samples.  
The results also showed that using a fixed forgetting factor yielded better performance than using a variable forgetting factor, which suggests that the effect of introducing a forgetting factor is small as the prediction length is limited. 
Figure~\ref{1988oracletime} shows the reference-signal waveform and the frame-wise NPR for $a=128$.
Overall, SP-FxLMS\_F+P achieved the highest speech suppression performance.

\begin{figure}[t!]
\centering
\includegraphics[width=0.90\linewidth]{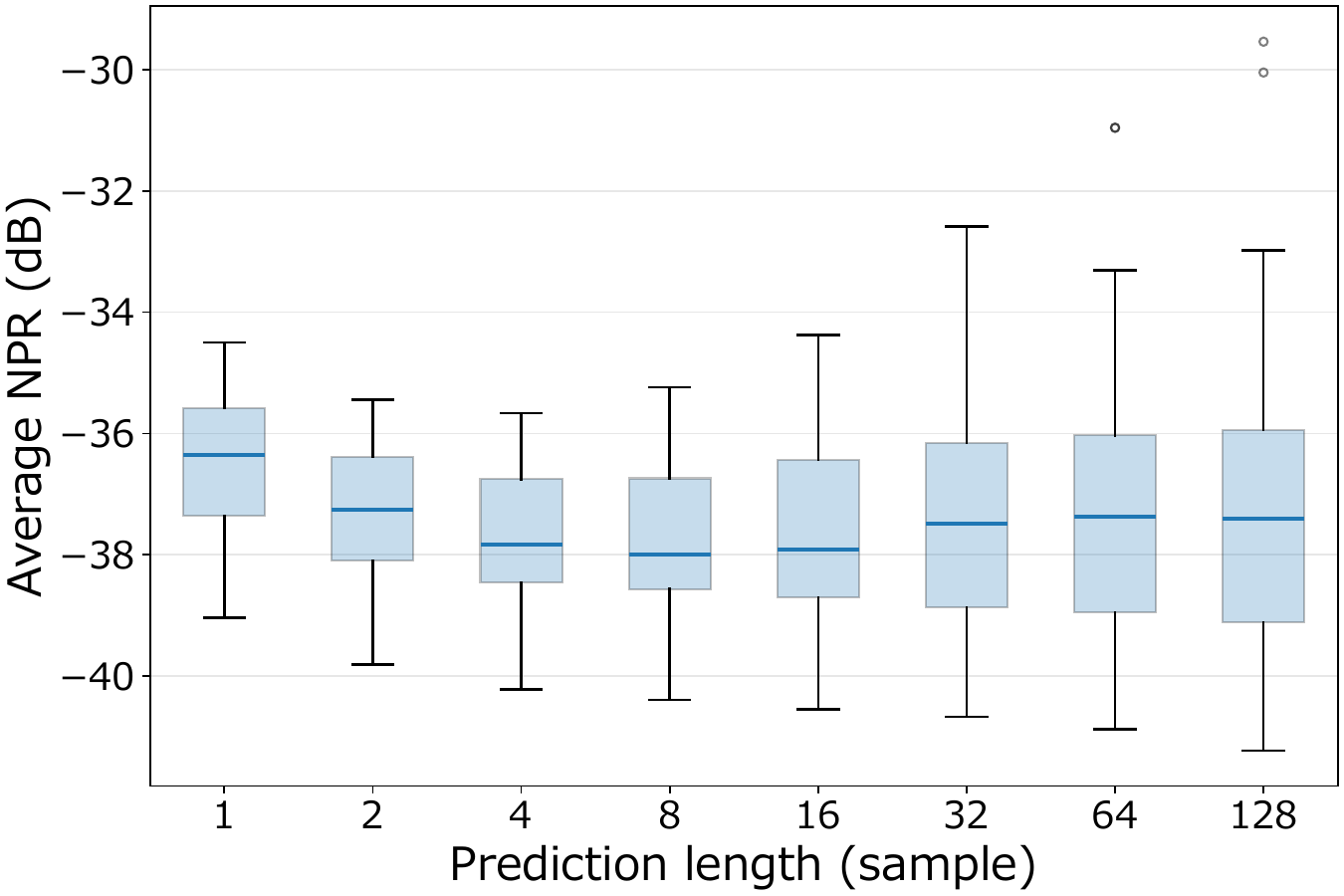}
\caption{
Boxplot of average $\mathrm{NPR}$ of SP-FxLMS\_F for 19 speakers with respect to the prediction length.
}
\label{multispeaker}
\end{figure}

Figure~\ref{multispeaker} shows the boxplot of the average NPR for 19 speakers with respect to the prediction length up to 128. 
Whereas the average NPR somewhat improved for prediction lengths of up to 8 samples, it becomes highly dependent on the speaker, particularly for prediction lengths of 32 samples or more. 

\begin{figure}[t!]
\centering
\includegraphics[width=0.90\linewidth]{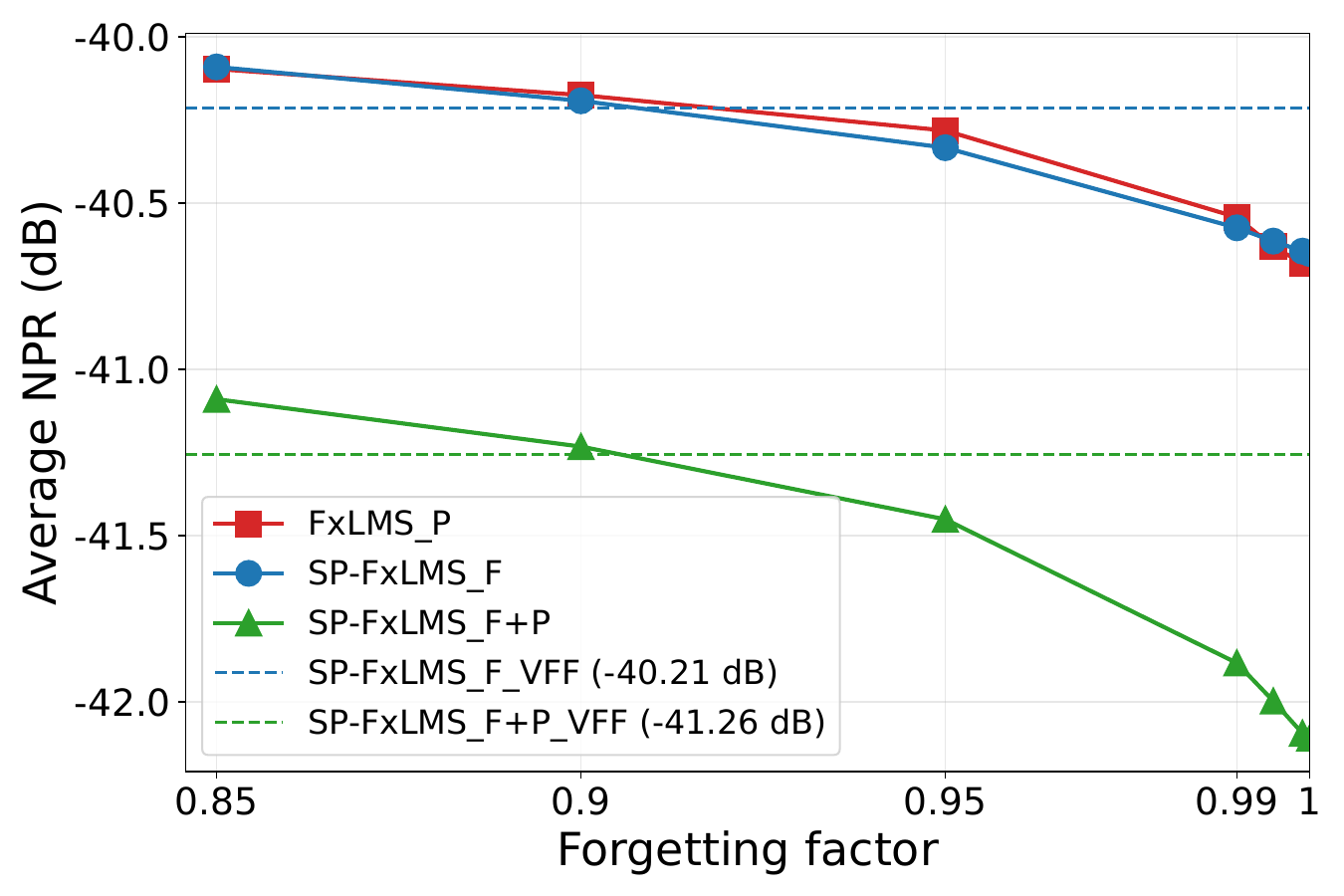}
\caption{
Average $\mathrm{NPR}$ for the prediction length $a=128$ with respect to the forgetting factor $\lambda$.
}
\label{1988oraclelamda}
\end{figure}

Figure~\ref{1988oraclelamda} shows the effect of $\lambda$ for $a=128$. All methods performed best at $\lambda=1$; thus, equal weighting was preferable to fixed or variable forgetting under this condition.

\subsection{Experiments using NN-based time-series prediction}

Next, the case in which the future signals predicted by using the MLP is investigated. We predicted only the reference signals, and the desired signal was generated by the given primary path. 

We used a single-hidden-layer MLP as a simple predictor to evaluate the effect of predicted gradients rather than to optimize the prediction architecture. We fixed the input length and prediction length as $N=512$ and $a=128$, respectively, and used 1024 hidden units with ReLU activation. The model was trained for 50 epochs using Adam~\cite{adam}, a learning rate of $10^{-3}$, and a batch size of $8$.
The training data were the speech signals of chapter ID \texttt{147956} and \texttt{148538}, and the test data were that of chapter ID \texttt{24833}, taken from \texttt{speaker 1988}. 

\begin{figure}[t!]
\centering
\includegraphics[width=0.85\linewidth]{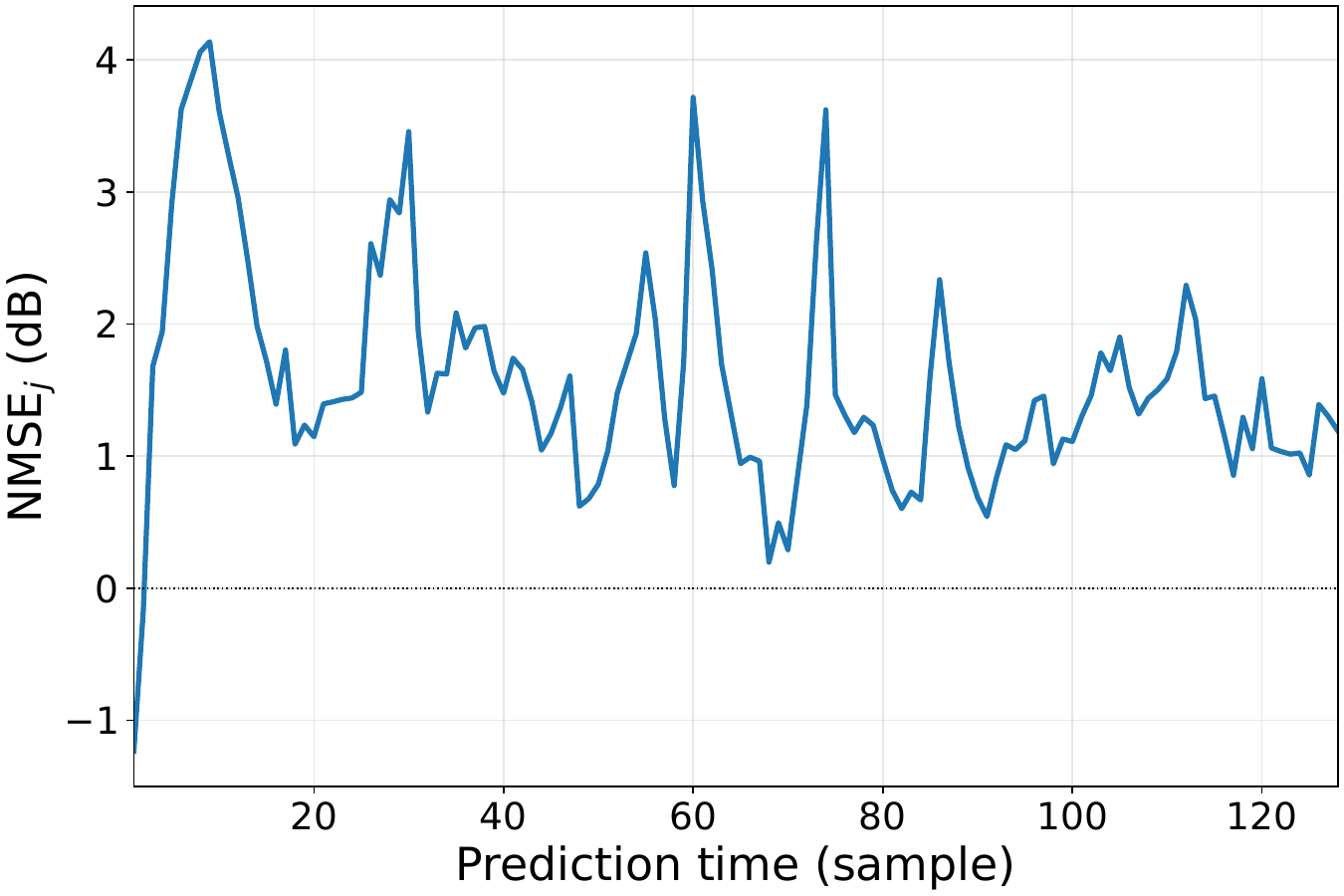}
\caption{
$\mathrm{NMSE}_j$ of the NN-based speech prediction with respect to the prediction time $j$. 
}
\label{predacc}
\end{figure}

The prediction accuracy for the $j$-sample prediction is evaluated by the normalized mean squared error (NMSE) as
\begin{equation}
\mathrm{NMSE}_j=10 \log_{10}\left(
\frac{\sum_k (x_{k+j}-\hat{x}_{k+j})^2}
{\sum_k x_{k+j}^2 }\right).
\end{equation}
The Pearson correlation coefficient is calculated between the entire true and predicted samples.
\begin{equation}
r=
\frac{\sum_{k,j}(x_{k+j}-\bar{x})(\hat{x}_{k+j}-\bar{\hat{x}})}
{\sqrt{\sum_{k,j}(x_{k+j}-\bar{x})^{2}
\sum_{k,j}(\hat{x}_{k+j}-\bar{\hat{x}})^{2}}},
\end{equation}
where \(\bar{x}\) and \(\bar{\hat{x}}\) denote the means of the true and predicted samples, respectively.
Figure~\ref{predacc} shows $\mathrm{NMSE}_j$ for $1 \le j \le 128$. 
The average NMSE and Pearson correlation coefficient were $1.67~\mathrm{dB}$ and $0.287$, respectively.
The MLP has $656,512$ parameters and requires $655,360$ multiply--accumulate operations per inference. Its average inference time on a CPU (Intel Core Ultra 7 155H) was $1.19 \times 10^{-4}~\mathrm{s}$.
Since this is longer than the sampling interval of $6.25 \times 10^{-5}~\mathrm{s}$, further implementation adjustments will be necessary for real-time processing. 

\begin{figure}[t!]
\centering
\includegraphics[width=0.90\linewidth]{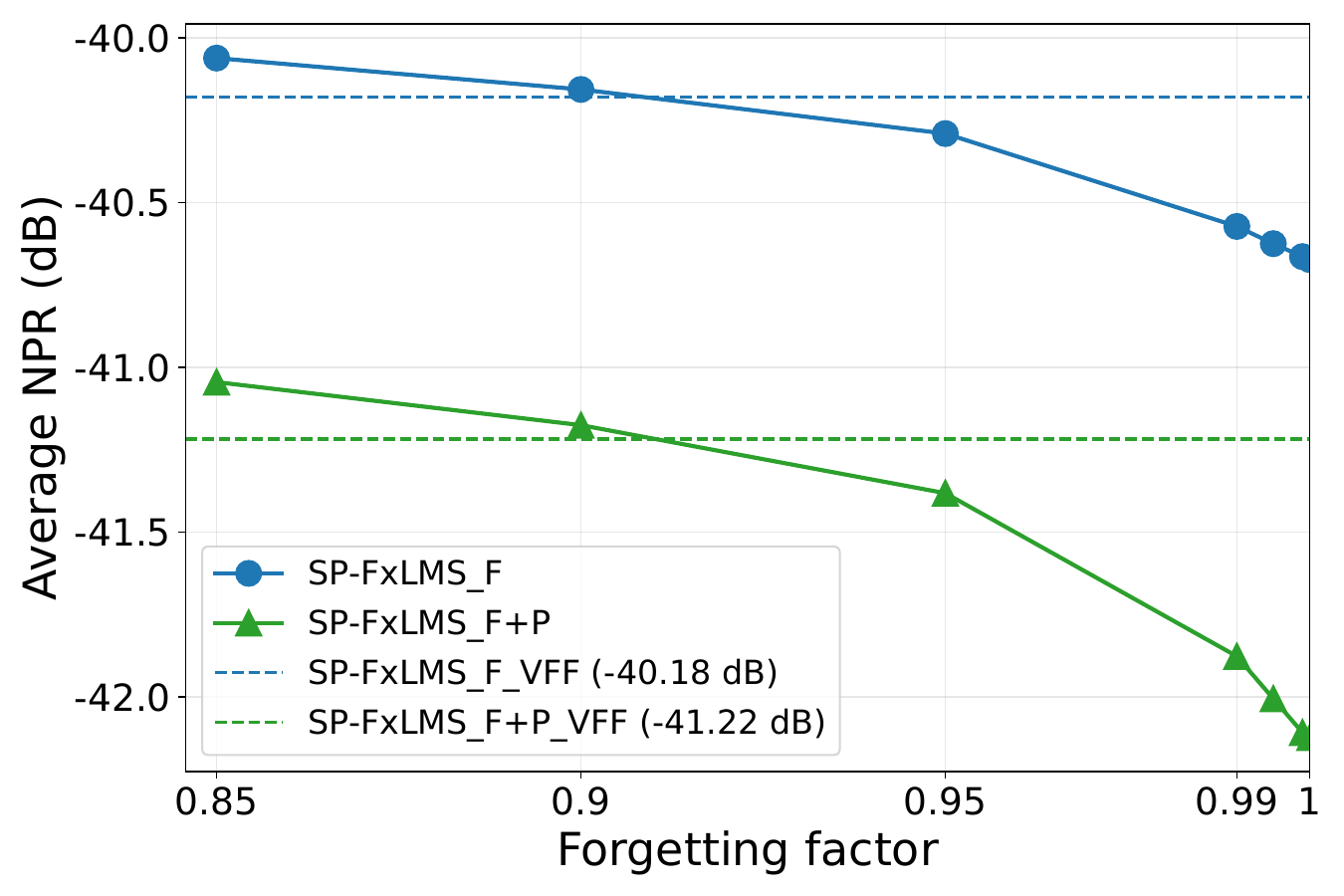}
\caption{
Average $\mathrm{NPR}$ for the prediction length $a=128$ with respect to the forgetting factor $\lambda$ when using the NN-based speech prediction.
}
\label{1988mlplamda}
\end{figure}

The effect of the forgetting factor when using the NN-based prediction is shown in Fig.~\ref{1988mlplamda}. 
As with the case using true predicted signals, the largest amount of speech suppression was achieved when the forgetting factor was 1. 
Even when the predicted signal contains errors, using equal weights for all predicted samples was effective when the prediction length was set to $a=128$.

\begin{figure}[t!]
\centering
\includegraphics[width=0.90\linewidth]{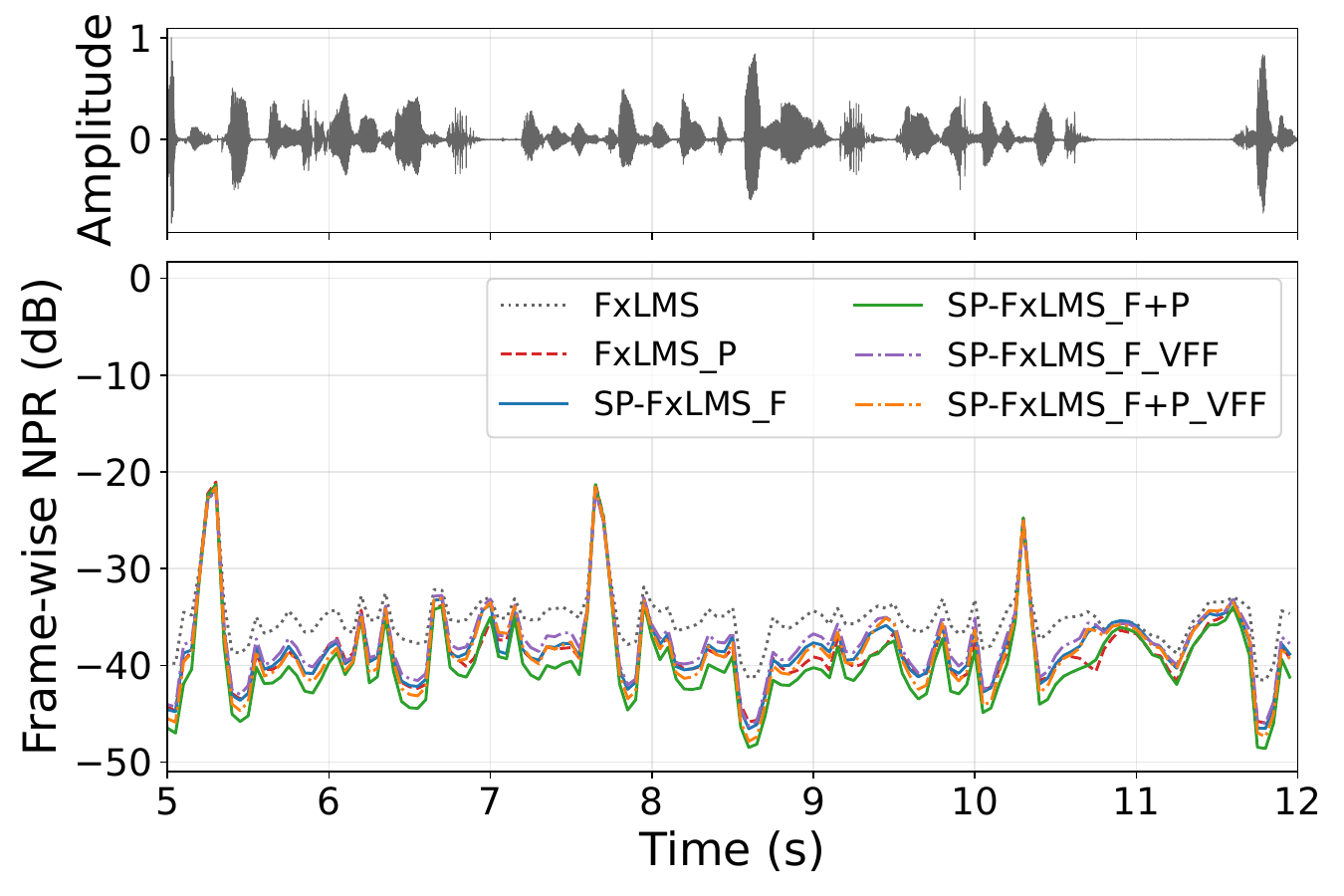}
\caption{
Waveform of the reference signal and frame-wise $\mathrm{NPR}$ when using the NN-based predicted signals for the prediction length $a=128$. 
}
\label{1988mlptime}
\end{figure}

\begin{table}[t]
\centering
\caption{Average $\mathrm{NPR}$ (\MakeLowercase{d}B) for $a=128$ when using NN-based time series prediction.}
\label{tab:npr}
\begin{tabular}{lc}
\hline
Method & Average $\mathrm{NPR}$ (dB) \\
\hline
FxLMS            & $-37.09$ \\
FxLMS\_P         & $-40.67$ \\
SP-FxLMS\_F      & $-40.67$ \\
SP-FxLMS\_F+P    & $\mathbf{-42.12}$ \\
SP-FxLMS\_F\_VFF   & $-40.18$ \\
SP-FxLMS\_F+P\_VFF & $-41.22$ \\
\hline
\end{tabular}
\end{table} 

Figure~\ref{1988mlptime} shows the reference-signal waveform and the frame-wise $\mathrm{NPR}$ for $a=128$ when using the NN-based predicted signals. 
Table~\ref{tab:npr} summarizes the average $\mathrm{NPR}$ for $a=128$.
FxLMS\_P and SP-FxLMS\_F achieved the same average NPR, indicating comparable benefits from past and predicted-future gradients under this condition. Combining both gradients provided an additional $1.45~\mathrm{dB}$ improvement over FxLMS\_P. The oracle and MLP predictions yielded nearly identical NPR values under this condition.

\section{Conclusion}

We proposed a time-series-prediction-based active speech suppression method. 
In the framework of the feedforward FxLMS algorithm, the update value of the linear control filter is calculated based on future, current, and past signals. 
The future signal is assumed to be predicted by NNs. 
Fixed and variable forgetting factors are also introduced in the filter update. 
In the numerical experiments, the proposed SP-FxLMS algorithm, which uses past, current, and future signals, achieved a $5.03~\mathrm{dB}$ improvement in noise reduction compared to the FxLMS algorithm, which only uses the current signal. This was also $1.45~\mathrm{dB}$ lower than that using the current and past signals. 
No significant difference in noise reduction was observed when using the true predicted signal or the signal predicted by the NN, although the average NMSE of the speech time-series prediction was $1.67~\mathrm{dB}$.

\section*{Acknowledgment}
This work was supported by JST FOREST Program, Grant Number JPMJFR216M.

\printbibliography

\end{document}